\documentstyle[12pt]{article}
\begin{document}
\baselineskip=20 pt
\def\l{\lambda}
\def\L{\Lambda}
\def\b{\beta}
\def\mphi{m_{\phi}}
\def\hphi{\hat{\phi}}
\def\vphi{\langle \phi \rangle}
\def\etamunu{\eta^{\mu\nu}}
\def\dmul{\partial_{\mu}}
\def\dnul{\partial_{\nu}}

\begin{center}
{\large\bf
  Implications of a light radion on the RG evolution of \\

\vskip 0.04in

  higgs self coupling in the  Randall-Sundrum model}  
 
\end{center}

\vskip 10pT
\begin{center}
{\large\sl \bf{Prasanta Das}~\footnote{E-mail: pdas@iitk.ac.in} 
}
\vskip  5pT
{\rm
Department of Physics, Indian Institute of Technology, \\
Kanpur 208 016, India.} \\
\end{center}


\begin{center}
{\large\sl  \bf{Uma Mahanta}~\footnote{E-mail:mahanta@mri.ernet.in}
}
\vskip 5pT
{\rm
Mehata Research Institute, \\
Chhatnag Road, Jhusi
Allahabad-211019, India .}\\
\end{center}

\centerline{\bf Abstarct}

In this paper we determine how the beta  funtion of the higgs self coupling
$\l $ at one loop order is modified by a light stabilized radion in the 
Randall-Sundrum model.
We then use the modified beta function to derive a lower bound on the radion
vev $\vphi$, both for perturbative and non-perturbative values of
$\l$ at the ultra violet cut off $\L$. The lower bound on $\vphi$ is 
obtained by demanding that the renormalized coupling $\l (\mu)$ at
$\mu $ =100 GeV be consistent with the present experimental bound
of 110 GeV on the higgs mass from LEP{$\Pi$} searches. We also show that
if $\l (\L )$ is 
sufficiently small then an upper bound on $\vphi$ can be 
determined by requiring that $\b (\l )$ be positive over the
relevant momentum range. 

\vfill\eject

\centerline{\large \bf Introduction}

Recently several attractive
proposals[1,2] based on theories in extra dimensions have been
put forward to explain the hierarchy problem. 
Among them the Randall-Sundrum model is particularly interesting
because it considers a five dimensional world based on the
following non-factorizable metric

$$ds^2= e^{-2k r_{c}\theta }{\eta_{\mu\nu}} dx^{\mu} dx^{\nu}- {r_c}^2 
d{\theta}^2
.\eqno(1)$$

Here $r_c$ measures the size of the extra dimensions which is an ${S^1/Z_2}$
orbifold. $x^{\mu }$ are the coordinates of the four
dimensional space-time. $-\pi\le \theta \le \pi$ is the coordinate
 of the extra dimension
with $\theta$ and $-\theta$ identified. k is a mass parameter of the order
of the fundamental five dimensional Planck mass M. 
Two 3 branes are placed at the orbifold
fixed points $\theta =0$ (hidden brane) and $\theta =\pi$ (visible
brane). Randall and Sundrum showed that any field on the visible brane
with a fundamental 
 mass parameter $m_0$ gets an effective mass $$m = m_0 e^{-k r_c \pi}$$
due to the exponential warp factor. Therefore
 for $k r_c \approx 14$ the electroweak scale
is generated from the Planck scale by the warp factor.

In the Randall-Sundrum model $r_{c}$ is the vacuum expectation 
value (vev) of a 
massless scalar field T(x). The modulus was therefore not stabilized
by some dynamics. In order to stabilize the modulus Goldberger and Wise[3] 
introduced a scalar field $\chi(x, \theta )$ in the bulk with interaction
potentials localised on the branes. This they showed could generate a
potential for $T(x)$ and stabilize the modulus at the right value 
($k r_{c} \approx 14$) needed for the hierarchy without any excessive
fine tuning of the parameters.

In the Randall-Sundrum model the SM fields are assumed to be localized
on the visible brane at $\theta =\pi$. However the SM action is modified
due to the exponential warp factor. Small fluctuations of the modulus
field from its vev gives rise to non-trivial couplings  of 
the modulus field with the SM fields. In this report we shall derive
the couplings of the radion to the higgs field up to quadratic order
in ${\hphi\over \vphi}$. Here $\hphi$ is a small fluctuation of the 
radion field from its vev and is given by $\phi =f e^{-k\pi T(x)}=
\vphi +\hphi$. $\vphi =f e^{-k\pi r_c}$ is the vev of $\phi$ and f
is a mass parameter of the order of M. We shall then detrmine the 
modification in the beta function for $\l$ to one loop due to
a light stabilized radion. The phenomenological implications
of the Randall-Sundrum model depends on two unknown parameters,
 the radion mass $\mphi$ and its vev $\vphi$. The requirement that the
interbrane seperation in the
Randall-Sundrum model be such so as to solve the hierarchy problem
implies that $\vphi$ must be of the order of a TeV.
Since the radion coupling to the SM fields is inversely proportional
to $\vphi$ the phenomenology of the RS model is expected to depend
quite sensitively on $\vphi$. In fact studies of radion phenomenology 
in the context 
of the RS model show that in order to be consistent with the
collider data $\vphi$ must be of the order of v (higgs vev) or greater[4].
In this paper we shall  use the RG equation for $\l$ in the RS model
to derive a lower bound on $\vphi$ for both perturbative and 
non-perturbative  values of $\l$ at the cut off scale $\L$. The lower bound
on $\vphi$ will be derived by demanding that the renormalized coupling 
$\l (\mu )$ at $\mu \approx $ 100 GeV should be consistent with the
present experimental bound of 110 GeV on the higgs mass.
We also show that if $\l (\L )$ is sufficiently small then it is possible 
to derive an upper bound on $\vphi$ by requiring that  $\beta(\lambda(\mu))$ 
must be positive for all $\mu \leq \Lambda$ .

\centerline{\bf Radion contribution to the RG equation for $\l $ }

The radion couplings to the higgs scalar is completely determined by
general covariance. The action for the higgs scalar in the 
Randall-Sundrum model can be written as

$$S=\int d^4x \sqrt {-g_v}[g_v^{\mu\nu}{1\over 2}\dmul h \dnul h -V(h)].
\eqno(2)$$

where $V(h)={1\over 2} \mu^2 h^2 +{\l \over 4} h^4$. h is a small
fluctuation of the higgs field from its classical vacuum v. 
In
abscence of graviton fluctuations we have
$$g_v^{\mu\nu}= e^{2k\pi T(x)}\eta^{\mu\nu}=({\phi\over f})^{-2}{\eta^{\mu\nu}}$$
$$\sqrt{-g_v} =({\phi\over f})^4 $$ where $$\phi = f e^{-k\pi T(x)}$$ 
Rescaling h
and v as $h\rightarrow {f\over \vphi}h$ and $v\rightarrow
{f\over \vphi} v$ we get

$$S=\int d^{4}x [({\phi\over \vphi})^2{1\over 2}{\eta^{\mu\nu}}\partial_{\mu} h \partial_{\nu} h-
({\phi\over \vphi})^4 V(h)].\eqno(3)$$
where
$$V(h)={\lambda\over 4} (h^4+4h^3v+4h^2v^2).\eqno(4)$$

The Feynman diagrams that give rise to the radion contribution to the
renormalization of the four higgs vertex  in the RS model are
shown in Fig 1. 

\vspace {-0.05in}

\begin{figure}[htb]
\begin{center}
\vspace*{1.2in}
      \relax\noindent\hskip -6.4in\relax{\includegraphics{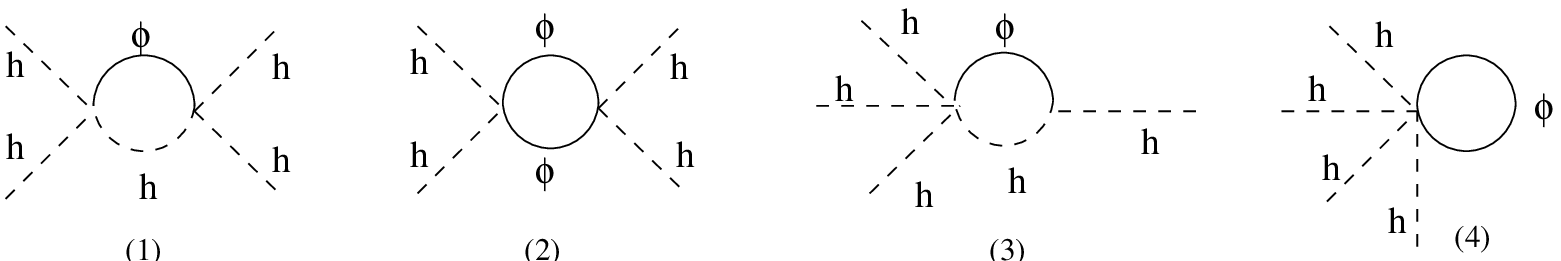}}
\end{center}
\end{figure}
\vspace*{0.05in}
\noindent {\bf Figure. 1}.
{\small { Feynman diagrams that give rise to the radion contribution
  to the vertex renormalization .}} 

It is clear 
from these diagrams that to evaluate them we need the couplings of one and
two radions to the higgs sector. Note first that the radion coupling 
to the kinetic energy term of the higgs boson will not contribute to the
renormalization of the vertex associated with the operator $h^4$. The reason 
being such couplings will give rise to operators 
involving derivatives of higgs field. Second the radion couplings to the
SM fields can be expressed as a power series expansion in ${1\over \vphi}$.
Hence naive dimensional analysis(NDA)[6] can be used to estimate the
ultraviolet (UV) cut off $\L $. Following the usual prescription of NDA  
we shall equate the cut-off $\L$ to $4 \pi \vphi$.
In general the the ratio ${\L\over \vphi}$ is expected to lie between
1 and 4$\pi$. However the estimates presented in this paper will
not change much as long as ${\L\over \vphi}$ lies in this range.
Further since perturbation theory is defined only about
a stable minimum we shall expand both h and $\phi$ about their respective
vevs. Evaluating the vertex renormalization diagrams explicitly with a cut
off $\L$ we find that the leading log terms of these diagrams are given by

$$\Gamma_1 = 6\l {288\l v^2\over 16\pi^2\vphi^2}\ln {\L^2\over \mu^2}.
\eqno(5a)$$.

$$\Gamma_2 =6\l {144\l v^4 \over 16\pi^2 \vphi^4} \ln{\L^2\over \mu^2}.
\eqno(5b)$$.

$$\Gamma_3 = 6\l {128\l v^2\over 16\pi^2\vphi^2}\ln {\L^2\over \mu^2}.
\eqno(5c)$$.

and

$$\Gamma_4 = -6\l {6\over 16\pi^2\vphi^2} [\L^2 -\mphi^2 
\ln {\L^2\over \mu^2}].
\eqno(5d)$$

Here $\mu$ is the renormalization mass scale. In the SM model the 
wavefunction renormalization constant of the higgs boson $Z_h$
is equal to one at one loop order even after the higgs field is 
shifted by its vev. However the radion coupling to the KE term 
of the higgs boson gives rise to a non-trivial wavefunction
renormalization of the higgs boson. Evaluating the radion mediated
self energy diagram (Fig.2) of the higgs boson,  

\begin{figure}[htb]
\begin{center}
\vspace*{0.5in}
      \relax\noindent\hskip -2.4in\relax{\includegraphics{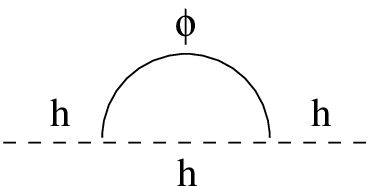}}
\end{center}
\end{figure}
\vspace*{-0.35in}
\noindent {\bf Figure. 2}.
{\small { Radion mediated self-energy diagram of the higgs boson.}} 

\vskip 10pt

we find that 
$Z_h= 1+{1\over 32\pi^2}{7m_h^2-\mphi^2 \over \vphi^2}\ln {\L^2\over
\mu^2}$.

Using the above vertex and wavefunction renormalizations induced by
a light radion it can be shown that the complete one loop beta
function for $\l$ in the RS model is given by 

\begin{eqnarray}
{\beta (\lambda )}  = \mu {d\lambda \over d\mu }={1\over 8\pi^2}[9\lambda^2 +
{402 \lambda^2 v^2\over \vphi^2}+ {144\lambda^2 v^4\over \vphi^4} + {7\lambda \mphi^2
\over \vphi^2}
+\lambda (6 g_y^2-{9\over 2} g^2-{3\over 2}g^{\prime 2})] 
\nonumber \\
+\frac{1}{8 \pi^2} [-6 g_y^4+{3\over 16} (g^4+{1\over 2}(g^2+g^{\prime 2})^2)] ~~ (6) \nonumber  
\end{eqnarray}

The purely SM contribution to $\beta(\lambda)$[7] can be obtained by
letting the expansion parameter $\vphi$ approach infinity.

\centerline{\bf Lower bound on radion vev }

For simplicity we shall first consider the higgs-radion system in isolation
 from the remaining fields. The beta function corresponding to this 
idealized situation can be obatined by setting $g_y=g=g^{\prime}=0$.
Such an approximation would be meaningful provided $\l (\mu )$
is much greater than the remaining couplings 
 over the entire momentum interval
of interest. Further for a light radion ( $\mphi \ll \vphi $ )
we can drop the term proportioanl to $\mphi^2$ from the expression
of $\beta (\l )$. The beta function for $\l$ then
 contains only quadratic terms in $\l$.
Solving the RG equation for $\l $ under the above
approximation we get 

\vspace*{-0.3in}

$$\l (\mu )= {\l (\L )\over 1+{\l (\L )\over 8\pi^2} (9+402 {v^2\over
\vphi^2}
+144 {v^4\over \vphi^4} )\ln {\L\over \mu}}\eqno(7)$$

In fig.3  we have plotted the renormalized coupling $\l (\mu )$ at $\mu $
=100 Gev against the radion vev $\vphi$ for $\l (\L )=\infty $ 
and $\l (\L )=$ e under the quadratic approximation to $\beta(\l)$ .

\begin{figure}[htb]
\begin{center}
\vspace*{3.1in}
      \relax\noindent\hskip -4.8in\relax{\includegraphics{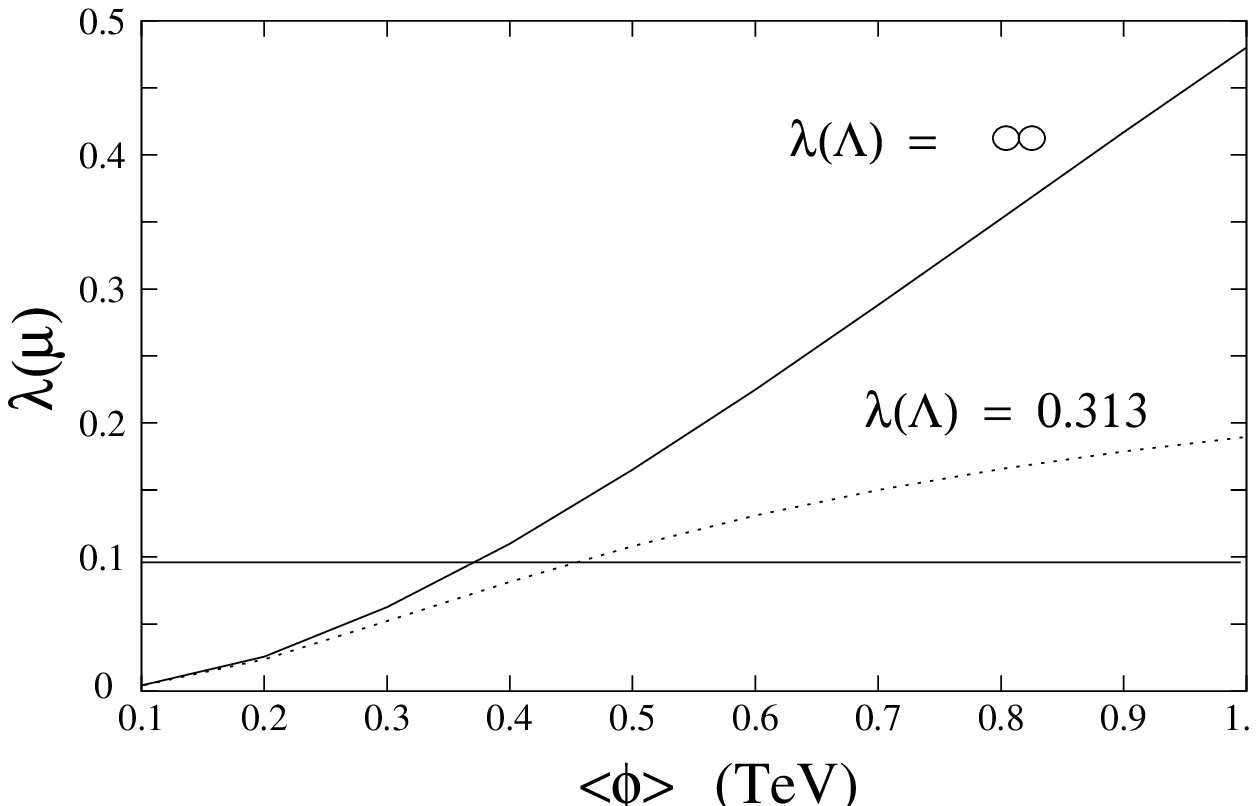}}
\end{center}
\end{figure}
\vspace*{-0.3in}
\noindent {\bf Figure.3}.
{\small{Showing the variation of $\lambda(\mu)$ at $\mu$ = 100 GeV 
with $\vphi$ when only the $O (\l^2)$ terms of $\b (\l)$ are kept.}} 

\vspace {0.1in}

We find that for $\l (\L ) =\infty $, in order that
$\l (\mu )$ at $\mu =100 $ GeV
 be greater than $0.099$ ( the value corresponding to the present
experimental bound on $m_h$) the radion
vev $\vphi$ must be greater than 378.2 GeV(solid 
curve).
The lower bound on $\vphi$ does not change much from this value
 as long as the value
of $\l (\L )$ remains non-perturbative i.e. $\l (\Lambda )\ge \sqrt {4\pi }$.
On the other hand if $\lambda$ lies in the perturbative regime e.g. if $\l
(\L )=e$ then in order that $\l (\mu )$
at $\mu =100$ GeV be consistent with the LEP II bound on the higgs mass,
the radion vev $\vphi$ must be greater than 468.4 GeV (dotted curve). 
The above results were obtained by keeping only the $O(\l ^2)$ terms in the
beta function for $\lambda $. If $\lambda (\L )$ is much greater 
than the remaining
couplings then clearly the evolution of $\lambda (\mu )$ towards low energies
will be determined mainly by the $O(\lambda ^2 )$ terms of $\b (\lambda )$.
However if $\l (\L )$ is small then the $O(\l ^2 )$ terms of $\b (\l )$
become smaller than the $O(\l )$ and $O(\l ^0)$ terms and the above
approximation breaks down. We have therefore considered the full
expression for $\b (\l )$ and determined the lower bound on $\vphi$
by demanding that $\l$ (100 GeV ) be consistent with the present 
experimental bound on $m_h$. 
By considering the full beta function for $\l $ and assuming for simplicity
that $g_y$, $g$ and $g^{\prime}$ do not scale with $\mu $ it can be shown
that the solution for $\l (\mu )$ is given by

\vspace {-0.25in}

$$\l (\mu )=\l_1 +{\l_1-\l_2\over {\l(\L )-\l_2\over \l(\L )-\l _1}
({\L\over \mu})^{a(\l_1-\l_2 )}-1}.\eqno(8a)$$

where $\l_1 ={-b+\sqrt {(b^2-4ac )}\over 2a}$ and 
$\l_2 ={-b-\sqrt {(b^2-4ac )}\over 2a}$

\vspace*{-0.2in}

$$a={1\over 8\pi^2}[9+402 {v^2\over \vphi^2}+144 {v^4\over \vphi^4}].
\eqno(8b)$$

\vspace*{-0.2in}

$$b={1\over 8\pi^2}[{7\mphi^2\over \vphi^2}+(6g_y^2-{9\over 2}g^2-{3\over 2}
g^{\prime 2})].\eqno(8c)$$

\vspace*{-0.2in}

and $$c={1\over 8\pi^2}[-6g_y^4+{3\over 16}(g^4+{1\over 2}(g^2+g^{\prime
2})^2)]
.\eqno(8d)$$




\newpage
\begin{figure}[htb]
\begin{center}
\vspace*{5.3in}
      \relax\noindent\hskip -8.7in\relax{\includegraphics{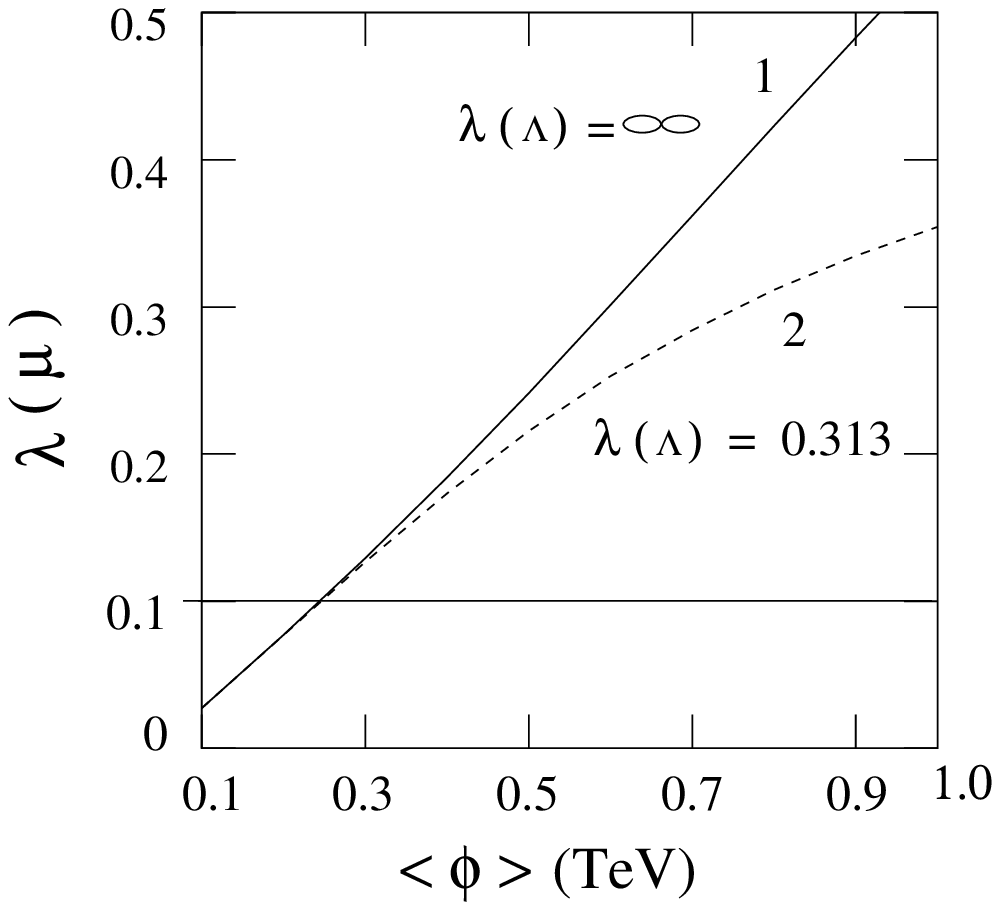}}
\end{center}
\end{figure}
\vspace*{-3.9in}
\noindent {\bf Figure.4}.
{\small {Showing the variation of $\lambda(\mu)$ at $\mu$ = 
100 GeV with $\vphi$ using the full expression for
$\beta(\lambda)$.}}



In Figure.4 we have plotted $\l (\mu )$ at $\mu $=100 GeV  against
different values of $\vphi $. The solid curve corresponds to the UV boundary
condition $\l (\L )=\infty $ and the dotted curve to $\l (\L )=e$.
Both plots were obtained with the following values of $g_y$, $g$ and
$g^{\prime}$: $g_y={{\sqrt 2}m_t\over v}=1.001$, $g={e\over \sin{\theta_w}}=
0.644$ and $g^{\prime}={e\over \cos{\theta_w}}= 0.356$. Further the radion 
mass $\mphi$ was assumed to be 50 Gev.
 From these two
plots we find that $\vphi$ must be greater than about $243~GeV$ so that 
$\l (\mu )$ at $\mu = $ 100 GeV is greater than 0.099($\approx{0.1}$). 
This estimate of lower bound on $\vphi$
 will not change much with $\mphi$ as long as
the radion is light and $\mphi$ lies in the few tens of Gev range.
We find that 
the lower bound on $\vphi$ obtained by using the complete expression
for $\beta(\lambda )$ does not depend at all on the UV boundary condition.
In fact fig. 4 shows that for $\vphi $ less than 250 Gev the
renormalized value of $\l (\mu )$ at low energies is governed
by the infrared properties of the theory and not the ultraviolet.

\centerline{\bf Upper bound on radion vev }

From the full expression of $\b (\l )$ it is clear that that if $\l$
is sufficiently small and $\vphi$ is large then $\b (\l )$ can become
negative due to the dominance of the $g_y^4$ term which is negative.
Hence for sufficiently small values of $\l (\L )$ a reasonable
 upper bound on $\vphi$ can be obtained by demanding that $\b (\l (\mu ))$
be positive for all $\mu\le \L$. This criterion will ensure that the 
RG evolution of $\l (\mu )$ from $\L$ towards low energies
exhibits infrared free behaviour. In particular $\b (\l (\L))$ must be
positive. 
At the crossover from infrared free behaviour to asymptotically
free behaviour the beta function vanishes and we get

\vspace*{-0.5in}

$$A(\l )x^2 +B(\l )x +C(\l )=0. \eqno(9a)$$
where 
$$A(\l ) = [9\l^2(\Lambda )+\l (\Lambda )(6g_y^2-{9\over 2}g^2-{3\over 2}g^{\prime 2})
-6g_y^4+{3\over 16}(g^4+{1\over 2}(g^2+g^{\prime 2})^2)].\eqno(9b)$$
$$B(\l )=402 \l^2(\Lambda )v^2+7\l (\Lambda)\mphi^2 .\eqno(9c)$$
$$C(\l )=144 \l^2(\L)v^4.\eqno(9d)$$
and $x^2=\vphi^2$.

Since $\vphi$ is real, x must be positive. Using this condition
it can
be shown that the physical root of the above equation is 
$x={-B(\l )-\sqrt {B^2(\l )-4A(\l )C(\l )}\over 2 A(\l )}$.

The upper bound on $\vphi$ for any given small value of $\l (\L )$ can be
determined from the above root. For example for $\l (\L )\approx e$ we
find that the radion vev must be greater than 806 Gev so that $\b(\l
(\L))$ is positive. In Figure 5 we have plotted the upper bound on $\vphi$
against $\l(\L )$. As expected the upper bound on $\vphi$ increases with
increasing $\l (\L )$. For $\l (\L )$ slightly greater than 0.6 both roots
become unphysical and no bound on $\vphi$ is obtained. The reason being
once $\l (\L )$ becomes sufficiently large $\b (\l )$ remains positive
irrespective of the value of $\vphi$. Note that the upper bound on $\vphi$
rises very sharply in the vicinity of this region. In fact our estimate 
for the upper bound becomes somewhat  unreliable here. 

\newpage

\begin{figure}[htb] 
\begin{center}
\vspace*{+5.3in}
\relax\noindent\hskip -8.8in\relax{\includegraphics{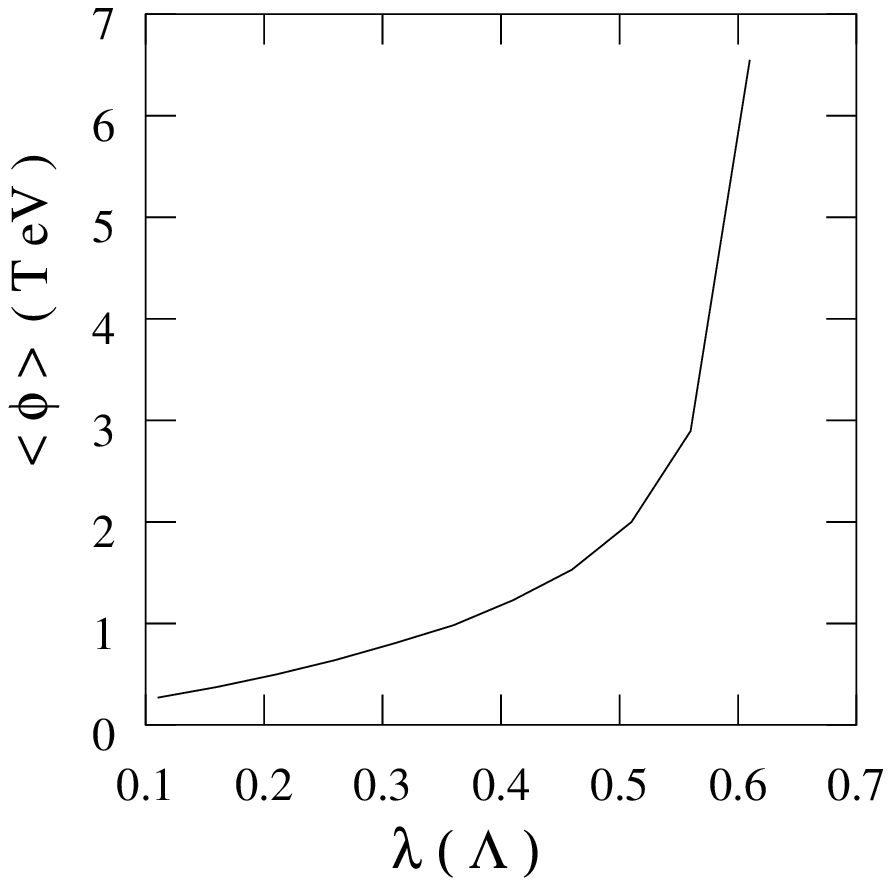}}
\end{center}
\end{figure} \noindent


\vspace*{-4.2in}

\begin{thebibliography}{92} 

\bibitem{N.A.Hamed} N. Arkani Hamed, S. Dimopoulos and G. Dvali,
Phys. Lett. {\bf B} 429, 263 (1998); I. Antoniadis, N. Arkani Hamed, S.
Dimopoulos and G.Dvali,Phys. Lett. {\bf B}463, 257 (1998).

\bibitem{L.Randall} L. Randall and R. Sundrum, Phys. Rev. Lett. 83, 3370
(1999).
\bibitem{W.D.Goldberger} W. D. Goldberger and M. B.
Wise, Phys. Rev. Lett. 83, 4922 (1999); C. Csaki, M. Graesser and 
L. Randall and J. Terning, Phys. Rev. {\bf D} 62, 045015 (2000);
W. D. Goldberger and M. B. Wise, Phys. Lett.
{\bf B} 475, 275 (2000).

\bibitem{U.Mahanta} U. Mahanta and S. Rakshit, Phys. Lett. {\bf B}480,
176 (2000); G. F. Giudice, R. Rattazzi and J. D. Wells,  Nucl. Phys. {\bf
B}595, 250 (2001); U. Mahanta and A. Datta, Phys. Lett. {\bf B}483, 196 (2000).

\medskip

For KK graviton phenomenology in the context of
 RS model see: H. Davoudiasl, J. A. Hewett
and T. G. Rizzo, Phys. Rev. Lett. 84, 2080 (2000); P. Das, S. Raychaudhuri
and S. Sarkar, JHEP 0007:050,2000 ;  D. K. Ghosh and S. Raychaudhuri,
Phys. Lett. {\bf B}495, 114 (2000).

\bibitem{C.Borean} C. Borean et al (ALEPH Collaboration), Phys. Lett. {\bf B}
495, 1 (2000); M. Acciari et al (L3 Collaboration), Phys. Lett. {\bf
B}495, 18 (2000).


\bibitem{H.Georgi} H. Georgi and A. Manohar, Nucl. Phys. {\bf B}234, 189
(1984); H. Georgi, Phys. Lett. {\bf B}298, 187 (1993); Z. Chacko, M. Luty
and E. Ponton, J. High Energy Physics 07, 036 (2000).

\bibitem{J.Gunion} J. Gunion, H. Haber, G. Kane and S. Dawson, The Higgs
Hunters Guide,(Addision-Wesley, Menlo Park, 1990).

\end{thebibliography}
\end{document}